\documentclass{article}
\usepackage{subfig}
\usepackage{graphicx}
\usepackage{amssymb}
\usepackage{authblk}
\usepackage[figuresright]{rotating}
\begin{document}
\title{AdiosStMan: Parallelizing Casacore Table Data System Using Adaptive IO System}

\author[1]{Ruonan Wang\thanks{jason.wang@icrar.org}}
\author[2]{Christopher Harris\thanks{chris.harris@pawsey.org.au}}
\author[1]{Andreas Wicenec\thanks{andreas.wicenec@icrar.org}}

\affil[1]{\it{International Centre for Radio Astronomy Research, The University of Western Australia, M468, 35 Stirling Hwy, Crawley, Western Australia 6009, Australia}}
\affil[2]{\it{Pawsey Supercomputing Centre, 26 Dick Perry Ave, Kensington, Western Australia 6151, Australia}}

\maketitle

\begin{abstract}
In this paper, we investigate the Casacore Table Data System (CTDS) used in the casacore and CASA libraries, and methods to parallelize it.
CTDS provides a storage manager plugin mechanism for third-party developers to design and implement their own CTDS storage managers.
Having this in mind, we looked into various storage backend techniques that can possibly enable parallel I/O for CTDS by implementing new storage managers.
After carrying on benchmarks showing the excellent parallel I/O throughput of the Adaptive IO System (ADIOS), we implemented an ADIOS based parallel CTDS storage manager.
We then applied the CASA MSTransform frequency split task to verify the ADIOS Storage Manager.
We also ran a series of performance tests to examine the I/O throughput in a massively parallel scenario.

\end{abstract}

\section{Introduction}
Modern radio astronomy is entering the era of big data. 
With the radio interferometry technique, the data production of a telescope scales quadratically with the number of antennas\cite{correlator}.
Next generation radio telescopes, such as the Square Kilometre Array (SKA), will consist of thousands of antennas.
Together with the prospective high frequency resolution and long baselines, it will produce petabytes of data per day \cite{skabaseline}.
The traditional method of processing radio astronomy data on desktops or workstations will have to be replaced by large scale clusters or supercomputers, in order to tackle the compute and data challenge.

However, most software packages and data formats traditionally used in radio astronomy data processing are poorly parallelized.
For instance, one of the most widely used software package for radio astronomy, the Common Astronomy Software Applications (CASA) library, used to only have OpenMP level parallelism that can collaboratively work on a maximum of a single node.
Introducing inter-node parallelism has long been planned, but it is only very recently that CASA has started to support MPI for some of its algorithms.
For data I/O, the Casacore Table Data System \cite{ctds} (CTDS) built-into the casacore library, used to only allow one process to write data to a table at a time.
It can be worked around by virtually concatenating multiply physical tables into a single logical table, so that from the logical table's perspective, it can be written in parallel.
However, there still has not been concrete solutions for parallel writing at the physical table level, which could be important for optimizing the I/O performance at the SKA scale.

Therefore, considering the data volume that next generation radio telescopes will generate, there is a possibility that the current data I/O technique becomes invalid.
It could be because of the enormous amount of files beyond the capability of any filesystems.
Or it could be because that data is too large to duplicate or move around, but can only be put into as few tables as possible where any applications can get access simultaneously.
This paper focuses on parallel data I/O issues of the Casacore Table Data System.
We will first look into the CTDS architecture, and then summarize the parallel I/O techniques that can potentially be used to parallelize CTDS without breaking the underlying architecture.
Following this, we will present our parallel CTDS storage manager design and implementation, as well as related testing results.

\section{Casacore Table Data System (CTDS) \& Storage Backends}

The casacore library \cite{casacore} is a set of common radio astronomy functions implemented in C++, originally derived from the AIPS++ library. 
As it forms the core algorithm component of the CASA (Common Astronomy Software Applications \cite{casapy}) library, it is currently one of the most widely used radio astronomy libraries.
Casacore has a well designed data I/O subsystem, namely the Casacore Table Data System (CTDS), which can handle terabyte-scale astronomy data efficiently. 
The Casacore Table Data System provides an abstract layer to data by defining the storage manager interface. 
Each column, or a group of columns, of a CASA table can be assigned with a storage manager that handles the actual I/O operations interacting with the storage backends.  
This essentially allows the substitution of the built-in storage managers with custom ones that are based on third-party storage backends. 

A CTDS table can be operated from multiple writers, but only in serial.
While one writer is writing data into a CTDS table, it locks the table and prevents others from writing until it is finished.
The built-in storage managers of casacore are also designed to comply with this serial writing mechanism.
As of now, in the newest version of casacore, 2.0.3, the lock mechanism can be disabled at compile time.
However, errors are still seen when multiple processes operate on a single table through this non-locking access.
There has been another workaround in CASA 4.5, which virtually concatenates multiple physical CTDS tables into one logical table, and thus enables parallel writing at the logical table layer.
At the physical table level, there still has not been concrete solutions for parallel writing.
Taking advantage of the storage manager interface, a promising direction would be to implement a custom storage manager based on a parallel storage backend and enabling parallelism at the physical table layer.

There are several categories of parallel storage backends that potentially work with CTDS, including filesystem based data formats, databases, database engines, and distributed object stores. 
Traditionally, filesystem based data formats have been the most effective way of dealing with scientific data. 
This is because they usually target a particular science scenario, or an abstract group of science scenarios.
A reflection of this is that these data formats are usually designed and optimized for numerical arrays, where other storage backend categories traditionally give very little consideration. 
Moreover, without involving overheads by reserving functionalities for index, query, security and so on, data files can be accessed for reading and writing quite efficiently, sometimes approaching the theoretical I/O bandwidth of storage hardware.
Some good examples of filesystem based data formats are FITS (Flexible Image Transport System \cite{fits}), HDF5 (Hierarchical Data Format Version 5 \cite{hdf5}) and ADIOS (Adaptive IO System \cite{adios}).
In these data formats, FITS has been most commonly used in radio astronomy.
However, there has long been a debate on how far FITS can still go since it does not support parallel I/O \cite{fitsfuture} \cite{replaceuvfits}.
HDF5 and ADIOS are more suitable solutions for future systems that desire a parallel storage backend.
Especially, ADIOS is essentially designed for large scale parallel IO, and in some cases, proved to be 1000 times faster than other parallel I/O libraries \cite{adios_1000}.

Over the last decade, another trend for managing scientific data is to use databases or a hybrid database and data file approach. 
This is because using the traditional filesystem based data formats, the management of metadata is usually relying on directory hierarchies and file names, which may not be easily scalable when it comes to petabyte scale data \cite{ms_data_report}.
However, very few traditional databases are actually optimized for, or even compatible with, the major form of scientific data, numeric arrays. 
Our preliminary investigation shows that one of the few databases that target scientific data, SciDB \cite{scidb}, performs one to two orders of magnitude slower than ADIOS or the casacore built-in storage manager, when given hundreds of large floating point arrays.
An improved solution is to directly use high performance database engines, or storage engines.
A good example is WiredTiger \cite{wiredtiger}, which has been recently adopted in MongoDB \cite{mongodb} as one of the optional underlying storage engines. 
This bypasses the database interface layer, and thus could possibly provide higher throughput than using a fully functioned database.

Similar ideas can also apply to filesystem based approaches.
Currently some parallel distributed filesystems, Lustre for instance, are essentially based on object stores, while modern object stores, such as Ceph \cite{ceph}, do also provide filesystem interfaces. 
This implies the possibility that higher throughput could also be expected by bypassing the filesystem interface and directly using object stores as the storage backend.
One difficult problem of this approach is that data centers do not always provide such low-level interfaces to end users, but rather usually a filesystem interface only.
In the meanwhile, setting up and maintaining such an object store on dedicated facilities at a sensible scale could be a considerable expense. 
This largely limits the universality of a pure object store based storage backend model. 

In this paper, we mainly focus on filesystem based data formats, as they proved to achieve relatively high throughputs at a reasonable cost in terms of both software development and hardware requirement. 
For the selection of the storage backend technique, we will summarize one of our representative preliminary investigations and present in the next section.

\section{Benchmarking HDF5 and ADIOS}

In this section, we will present some benchmark results comparing the parallel I/O throughput of the HDF5 and ADIOS libraries.
This benchmark is one of our preliminary investigations into suitable storage backends that has greatly impacted the final decision of the technique we used to implement the CTDS storage manager.

\subsection{Scenario}
\begin{figure}
\includegraphics[width=\textwidth]{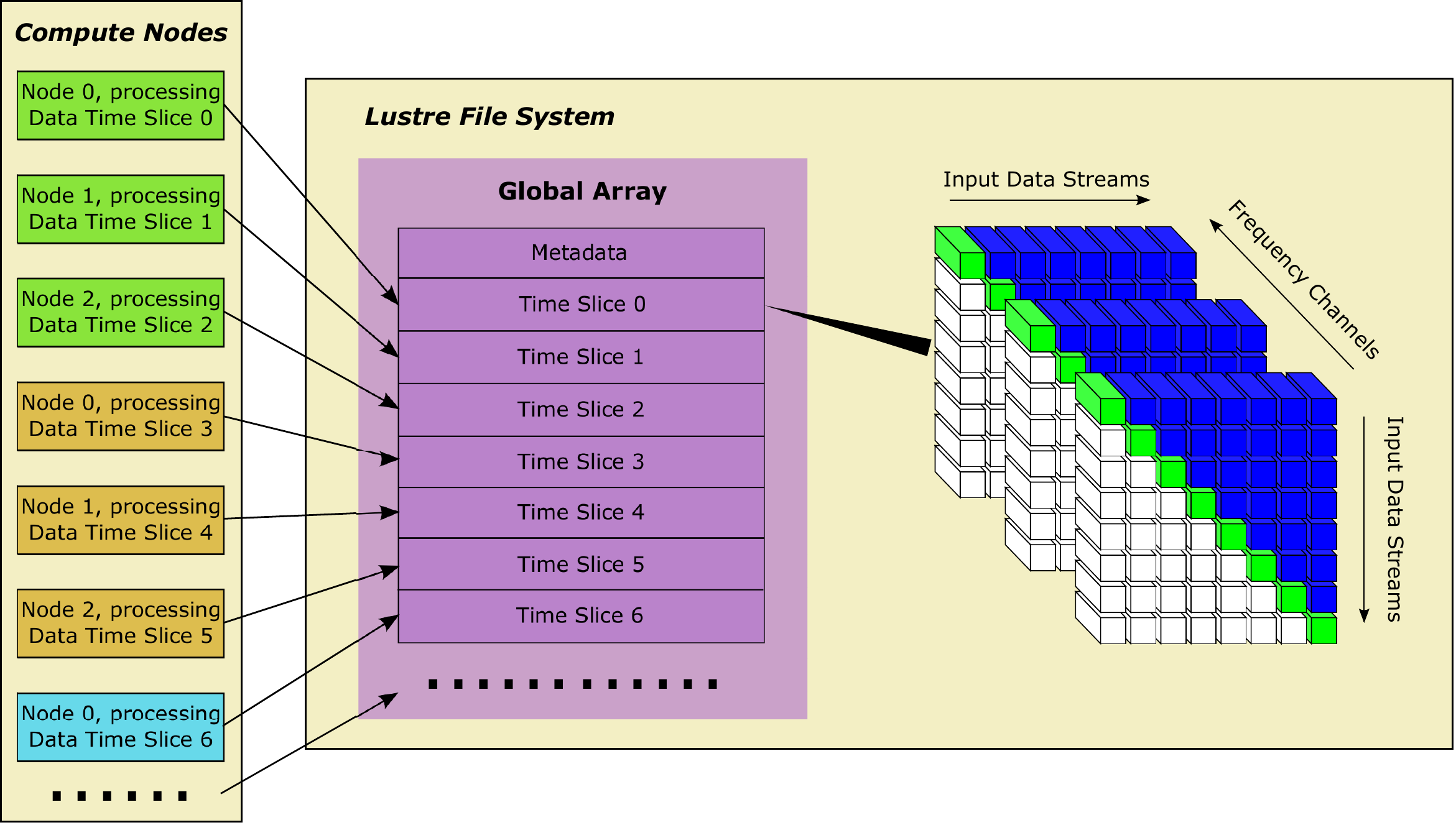}
\caption{Shown is the data output pattern of the time division multiplex correlator we used in the benchmark}
\label{fig:bench_flow}
\end{figure}

The benchmark code used in this section is a GPU cluster based radio astronomy signal correlator~\cite{correlator}.
It works in a time-division multiplex mode, and as a result, output data is not necessarily aligned in the desirable time sequence without specific data patterns and data flow logics involved.
To solve this problem, the I/O module of the correlator code uses pre-defined tables, where each row stores a time slice of output data, as shown in Figure~\ref{fig:bench_flow}.
In such a data paradigm, only asynchronous I/O is needed, which saves the overhead caused by blocking collective I/O operations.
For HDF5, this is done through pre-defining a dataset and then filling data in using hyperslabs.
Since no synchronous I/O is required, the serial HDF5 library is used instead of the parallel HDF5 compiled with MPI environment.
Similarly, for ADIOS, an ADIOS global array is pre-defined as the data container table, with each row subscribed to an ADIOS VAR for time slice data to fill in later.
In order to saturate the I/O module, the GPU compute module is switched off throughout the benchmark. 

\subsection{Testbed}
\label{section:testbed}
Two supercomputing systems at the Pawsey Supercomputing Centre, \textbf{Magnus} and \textbf{Fornax}, were used in this benchmark.
\textbf{Magnus} is a petascale Cray XC30 supercomputer, after a two-phase construction.
When this benchmark was conducted, it was at its first phase, consisting of 208 compute nodes. 
Each node contained two Intel Xeon E5-2670 CPUs, with the eight core Sandy Bridge architecture, and 64 GB of random access memory.
The nodes were interconnected by an Aries interconnect in a dragonfly topology, capable of 72Gbps of bandwidth per node.

In addition to the compute nodes, Magnus had 24 service nodes, which route traffic between the Aries interconnect and an Infiniband network.
The latter provides access to the Lustre version 2.2.0 filesystem, provided by a Cray Sonexion 1600.
This has two petabytes of storage via nine Scalable Storage Units (SSUs).
These SSUs have 8 OSTs, each using a 8+2 RAID 6 configuration.
The specification of each SSU has a 5 GB per second bandwidth from the IOR benchmark, and thus the expected peak bandwidth is 45 GB per second.

In terms of software on Magnus, the GCC version 4.7 compiler was used, along with the Cray MPICH version 6.1 library, HDF5 version 1.8.12 and ADIOS version 1.4.1.

The other system \textbf{Fornax} is an SGI machine designed for data intensive research, especially radio astronomy related data processing.
It consisted of 96 nodes, each containing two Intel Xeon X5650 CPUs, an NVIDIA Tesla C2075 GPU and 72 gigabytes of system memory.
The Intel 5520 Chipset is used in the computing node architecture, which enables the NVIDIA Tesla C2075 GPU to work on an x16 PCI-E slot and two QLogic Infiniband IBA 7322 QDR cards to run on two x8 PCI-E slots.

The back-end of Fornax's Lustre system is a SGI Infinite S16k, which is a re-badged DDN SFA 10k, consisting of 8 Object Storage Servers (OSSs) and 44 Object Storage Targets (OSTs), of which 32 are assigned to the scratch file system used in this testing.
Each of the OSSs has dual 4x QDR Infiniband connections to the switch connecting compute nodes, and the OSTs are connected to the OSSs via 8 4x QDR Infiniband connections.
Each OST consists of 10 Hitachi Deskstar 7K2000 hard drives arranged into a 8+2 RAID 6 configuration.
Operational testing using the ost\_survey Lustre benchmark achieved a mean bandwidth of 343 MB per second, and thus the expected bandwidth is approximately 11 GB per second. 

The software used on Fornax included the GCC version 4.4 compiler, OpenMPI version 1.6.3, HDF5 version 1.8.12 and ADIOS version 1.4.1.

\begin{figure}[!bh]
\begin{center}
\subfloat{\includegraphics[width=0.46\textwidth]{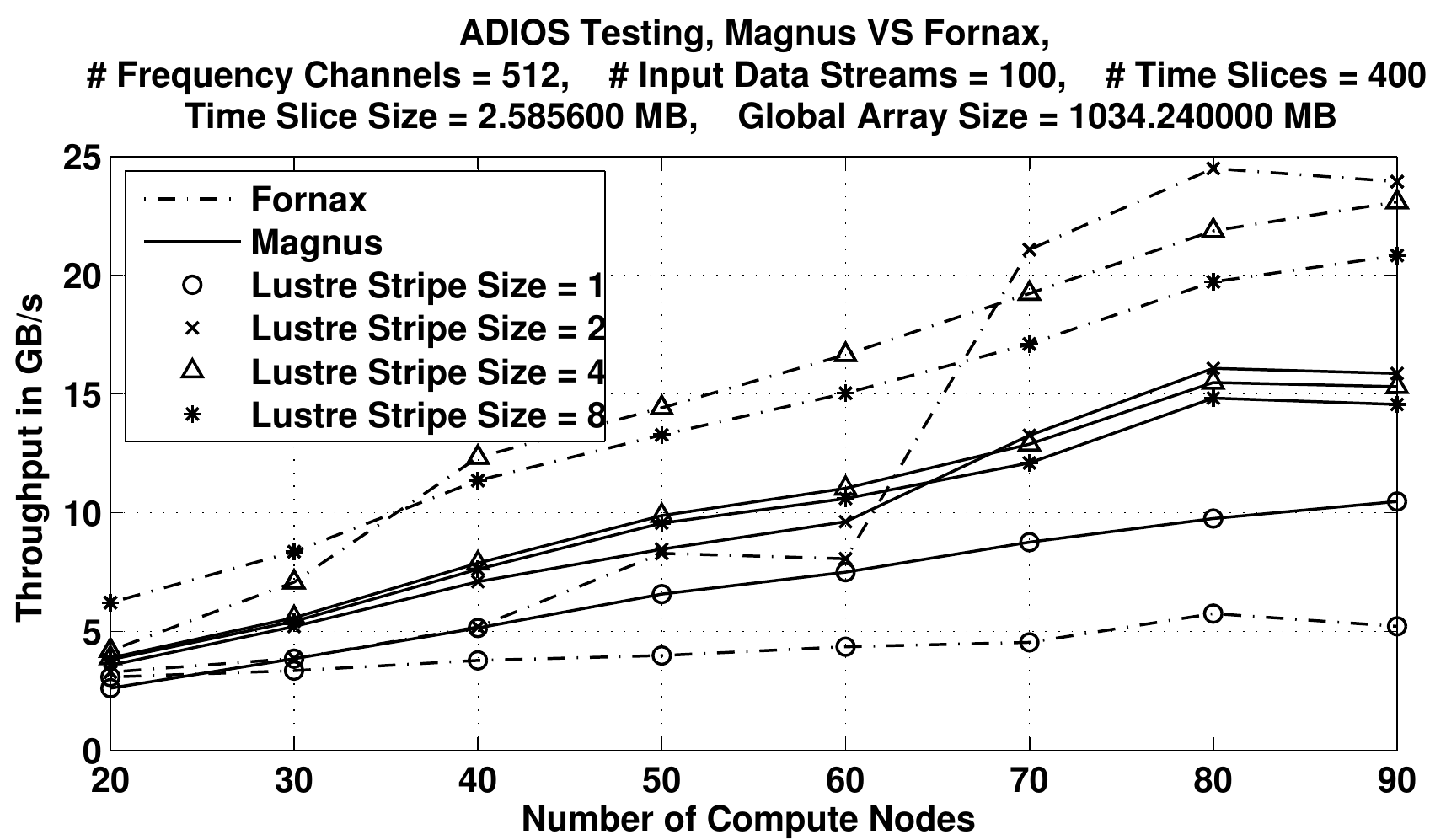}}
\subfloat{\includegraphics[width=0.46\textwidth]{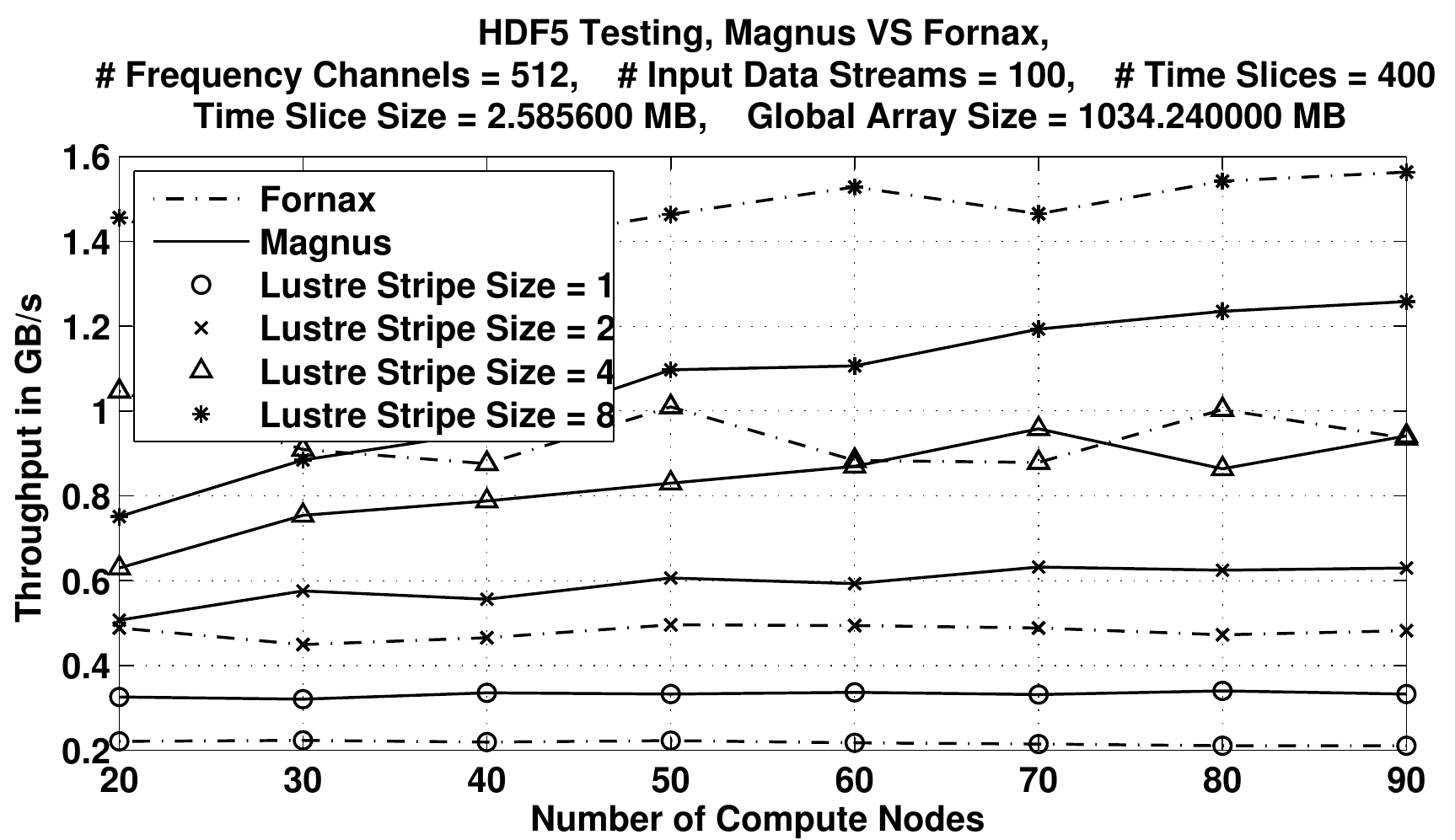}} \\
\subfloat{\includegraphics[width=0.46\textwidth]{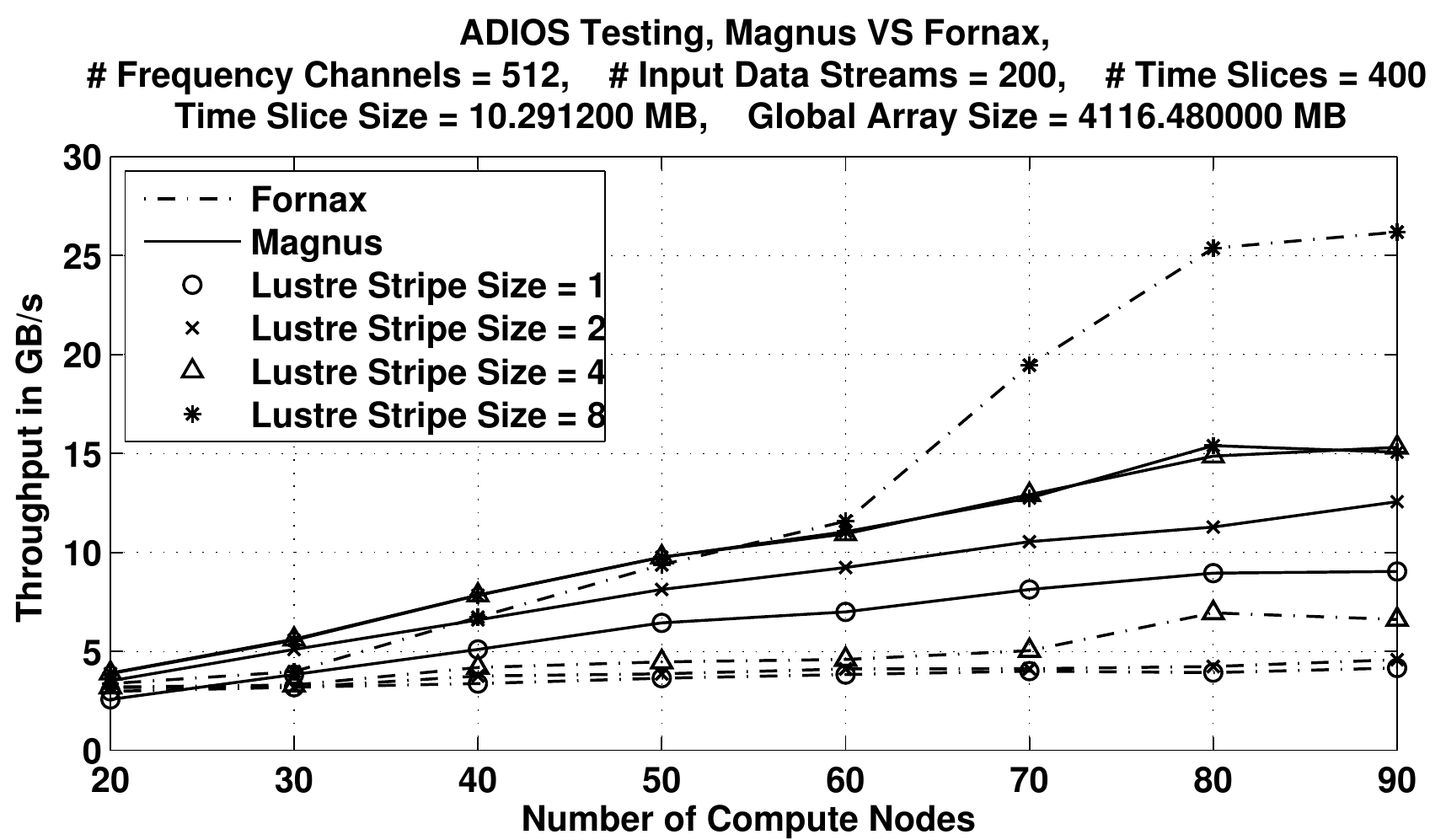}}
\subfloat{\includegraphics[width=0.46\textwidth]{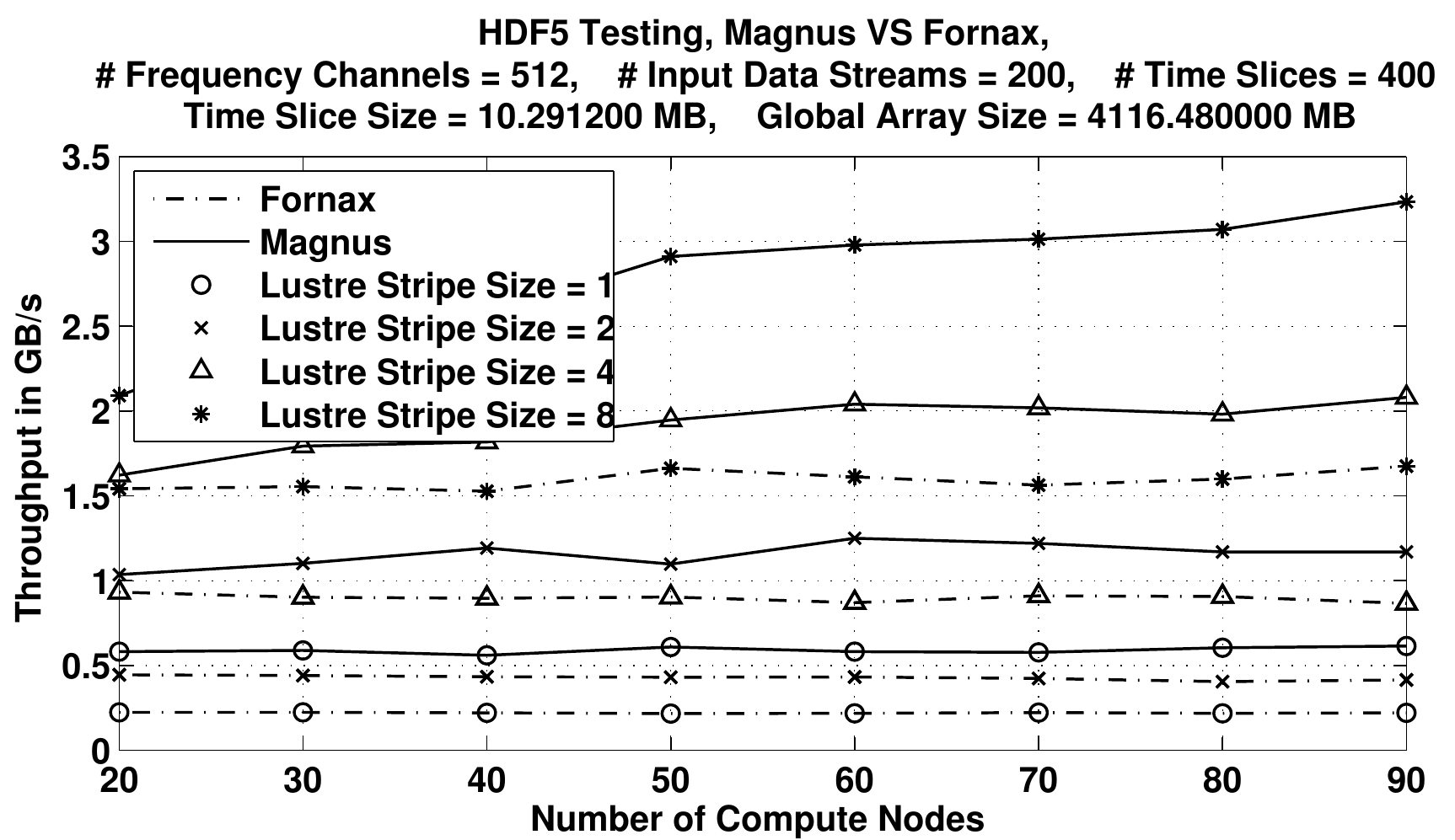}} \\
\subfloat{\includegraphics[width=0.46\textwidth]{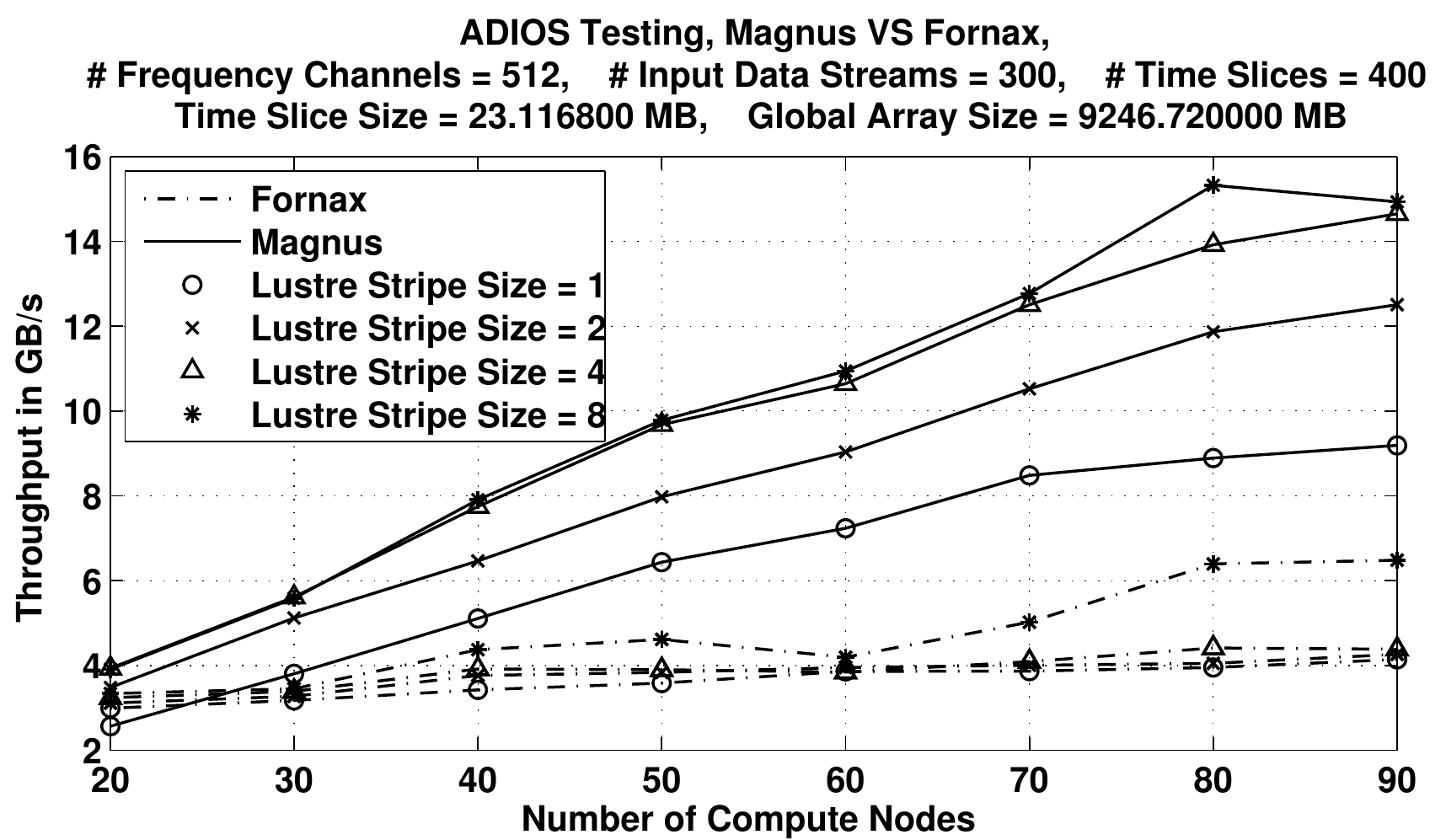}}
\subfloat{\includegraphics[width=0.46\textwidth]{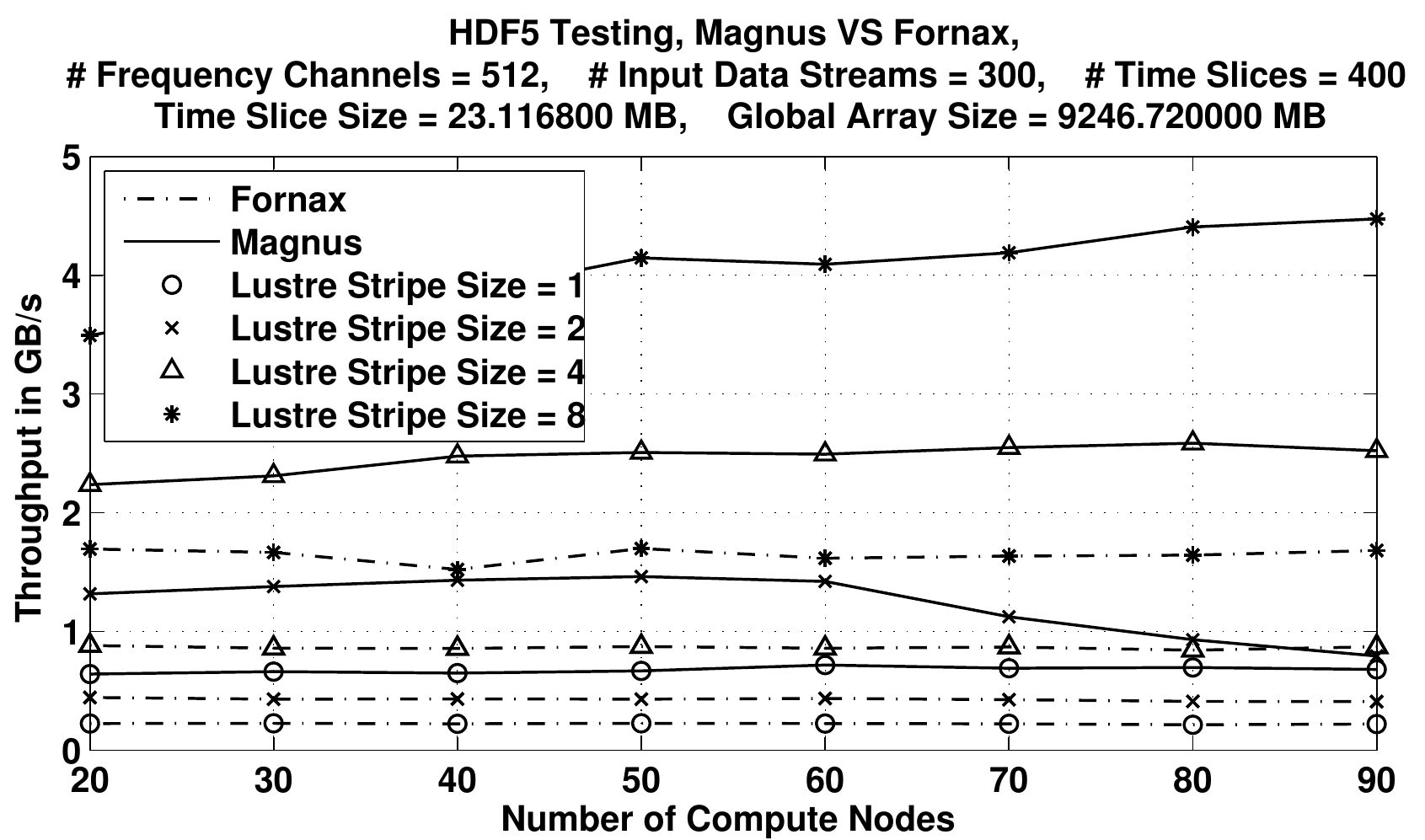}} \\
\subfloat{\includegraphics[width=0.46\textwidth]{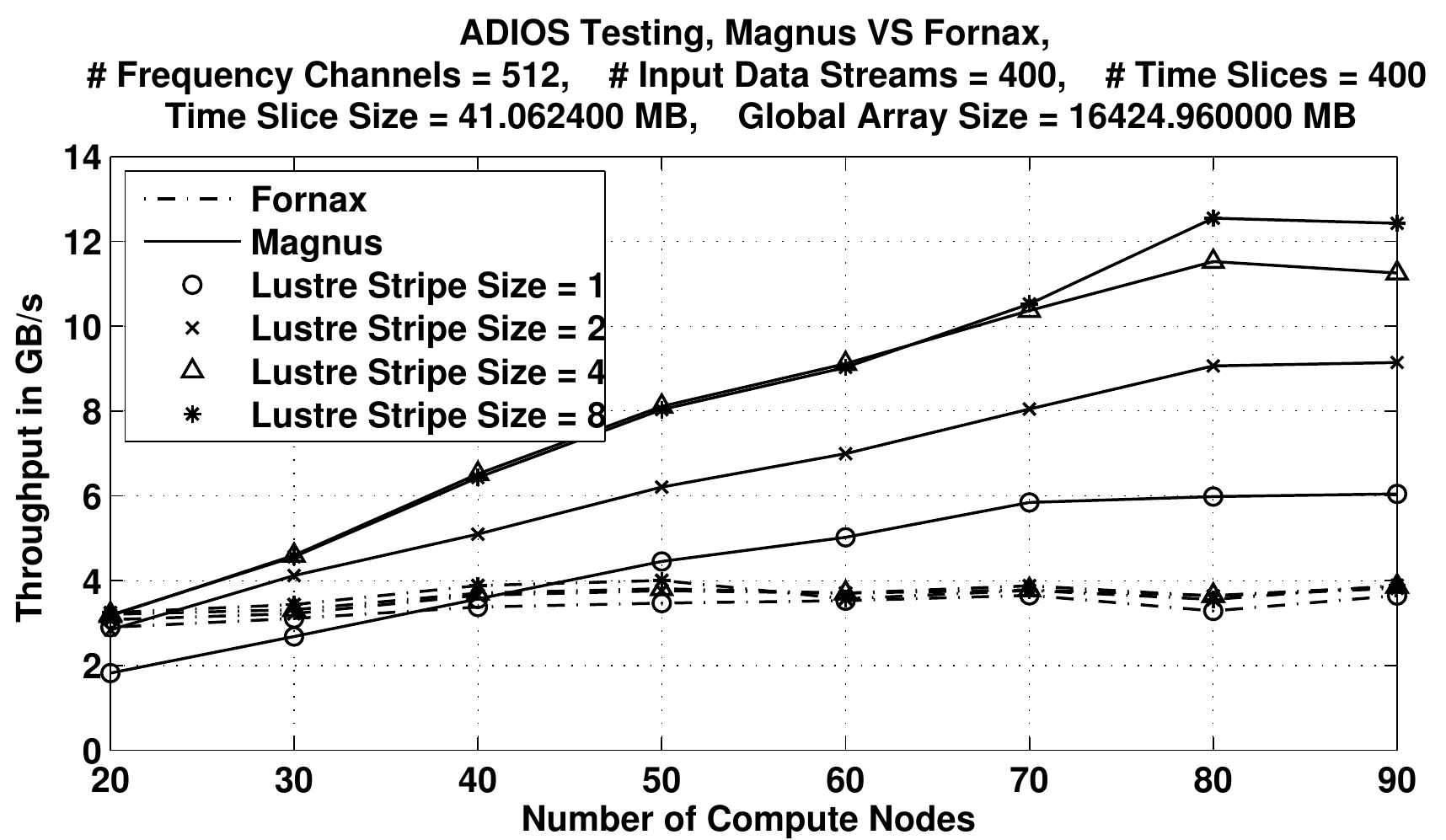}}
\subfloat{\includegraphics[width=0.46\textwidth]{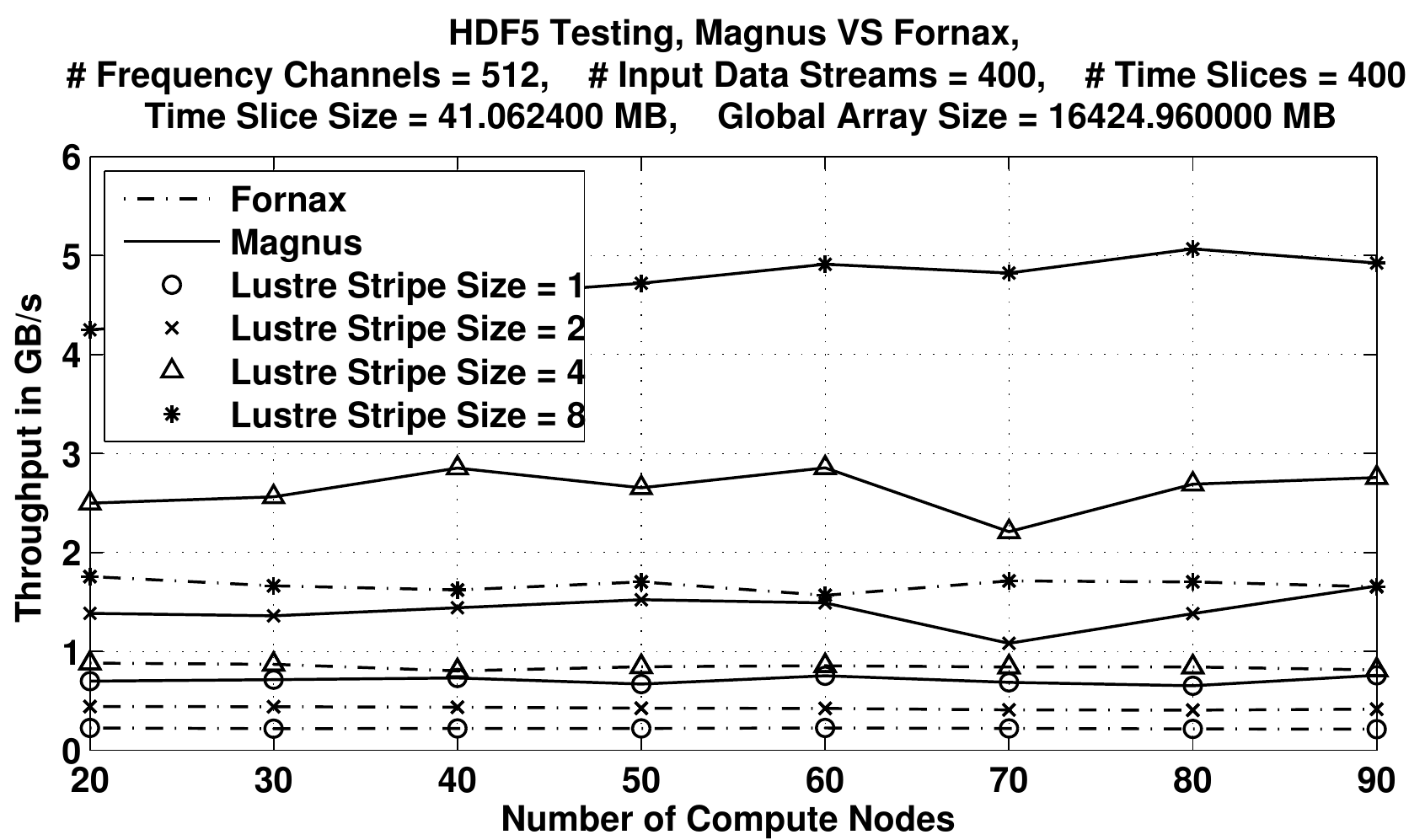}}
\caption{Shown is the ADIOS and HDF5 benchmark results obtained from the Magnus and the Fornax supercomputers at Pawsey Supercomputing Centre. The benchmark is configured to write a 1 GB to 16 GB global array from 20 to 90 compute nodes, each node launching a single MPI process, comparing ADIOS and HDF5 performance for various Lustre stripe sizes.}
\label{fig:bench}
\end{center}
\end{figure}

\subsection{Result}
Considering the scope of this paper, we only present a limited subset of the benchmark results in Figure \ref{fig:bench}, while readers are directed to our previous publication \cite{cug2014} for more detailed results and discussions. 
Within the parameter range of this subset, ADIOS performs 2 to 16 times faster than HDF5 depending on the global array size. 
Together with our other similar tests and benchmarks that are not presented due to space constraints, it would seem sensible to choose ADIOS as the storage backend to carry out the casacore storage manager experiment.

\section{ADIOS Storage Manager}

Based on the benchmark results presented in the last section, we then implemented a CTDS storage manager using ADIOS as its storage backend. 
Following the CTDS naming convention, the storage manager is named AdiosStMan, as a contraction of the ADIOS Storage Manager.
AdiosStMan is designed to be an alternative to any other CTDS storage managers while working in serial.
Any existing code using the CTDS system can work with AdiosStMan by simply changing the definition of the storage manager to AdiosStMan, provided that all CTDS features it uses are supported by AdiosStMan.
The CTDS features that the current version of AdiosStMan supports are given in Table \ref{tab:support_list}.

\begin{table}[h]
\caption{Shown is the list of CTDS features supported by the current version of AdiosStMan}
\label{tab:support_list}
\begin{center}
\begin{tabular}{ c | c }
\hline
Data types & All supported except String / String Array\\ \hline
Column Types & Scalar columns and direct array columns supported \\ \hline
Write & Supported \\ \hline
Rewrite & Not supported \\ \hline
Read & Supported \\ \hline
Add Rows & Not supported \\ \hline
Add Columns & Not supported \\ \hline 
\end{tabular}
\end{center}
\end{table}

\begin{figure}
\includegraphics[width=\textwidth]{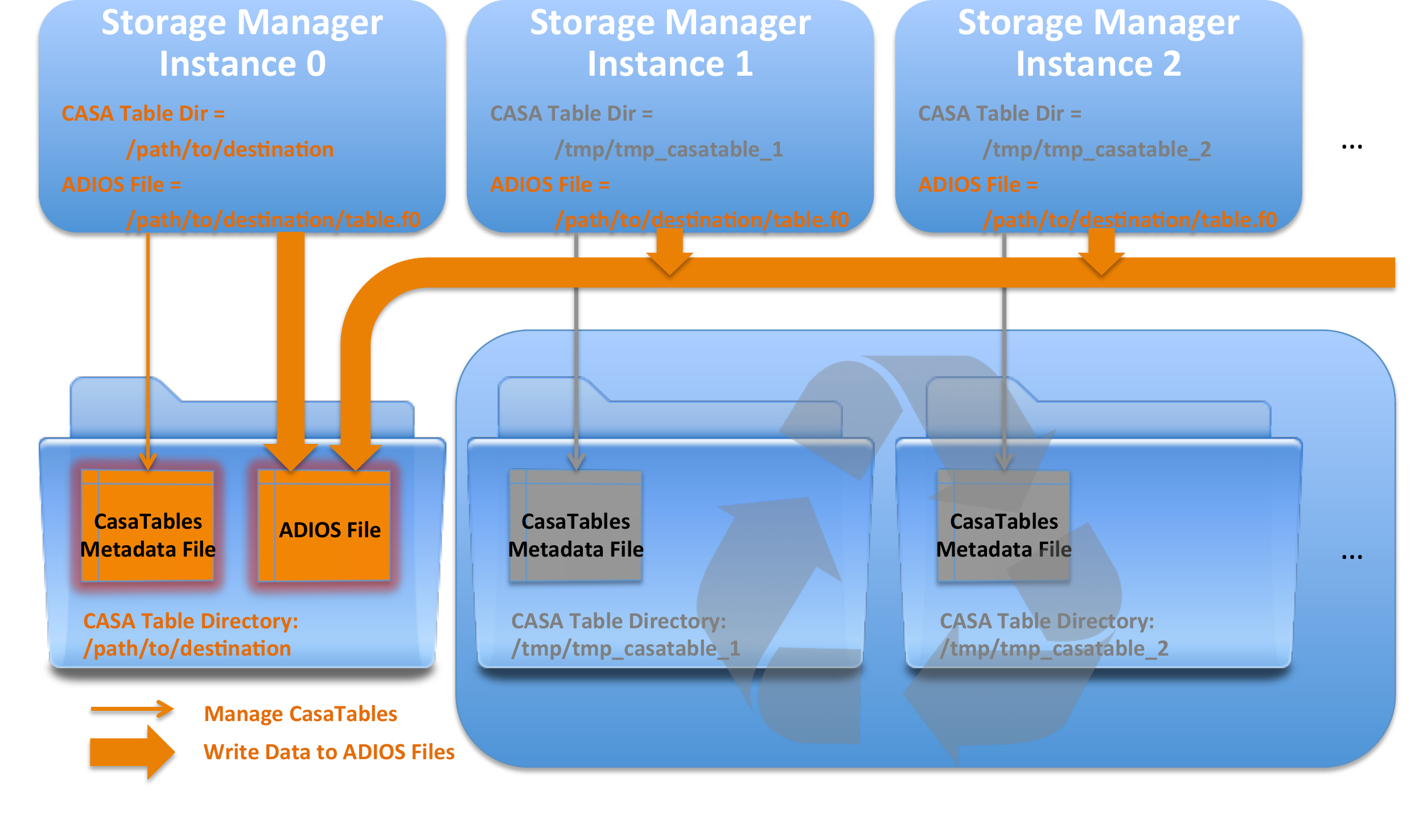}
\caption{Shown is the data flow of AdiosStMan working in the multi-process scenario.}
\label{fig:adiosstman}
\end{figure}

CTDS has a locking mechanism preventing multiple processes from accessing a physical table at the same time for writing.
To break through this limit and achieve parallel writing at the physical table level, AdiosStMan is designed in a way that when multiple processes access the same table, only the master process actually manipulates the table metadata file, while slave processes are fooled to write their table files into a temporary path.
A clearer illustration of this model is given in Figure \ref{fig:adiosstman}.
In the meantime, once AdiosStMan is instantiated and bound to a table from multiple processes, the master process broadcasts the path and name of the ADIOS file associated with the intended table metadata file to slave processes, for the AdiosStMan instances in slave processes to access.
The Casacore Table Data System then works as if every process is handling completely independent tables, although actually the data coming from all processes is put into the same ADIOS file that is associated with the table file of the masters process. 
Once all AdiosStMan instances are finalized, the temporary table files generated by slave processes can then be removed.
The ADIOS file together with the table file produced by the master process contains all necessary information to rebuild the entire table. 

From the programming paradigm aspect, AdiosStMan is built upon the MPI environment.
It supports MPI based multi-process code, but it is not thread safe and does not provide any multi-thread interfaces.
AdiosStMan is aware of being instantiated in a multi-process MPI code in both implicit and explicit cases. 
In the implicit case, no MPI related code has been executed at all by the time AdiosStMan is instantiated.
Then AdiosStMan will initialize MPI and pass any necessary configurations to the ADIOS library that relies on the MPI environment. 
When an AdiosStMan object is destructed, it finalizes MPI as well.
In the explicit case, if MPI has been already initialized when AdiosStMan is being instantiated, then AdiosStMan will carry on using the existing MPI configuration. 
To allow other parts of the code to continue using the MPI configuration, the AdiosStMan object will not finalize MPI while being destructed in this case.

With previous CTDS storage managers, when the casacore library is called from multiple processes, each process can only access its own dataset.
For next generation telescopes that involve a large number of processes in a data pipeline, the management of datasets and metadata would become a great expense.
Being the first parallel CTDS storage manager, AdiosStMan enables a type of use cases where the casacore library is called from multiple MPI processes, and these processes desire to dump data into a single dataset.
This provides a low-developing-cost solution to scale the existing data models orders of magnitude further towards the peta-scale goal for next generation radio telescopes.

\section{MeasurementSet Table Experiment}

\begin{figure}
\begin{center}
\includegraphics[width=0.9\textwidth]{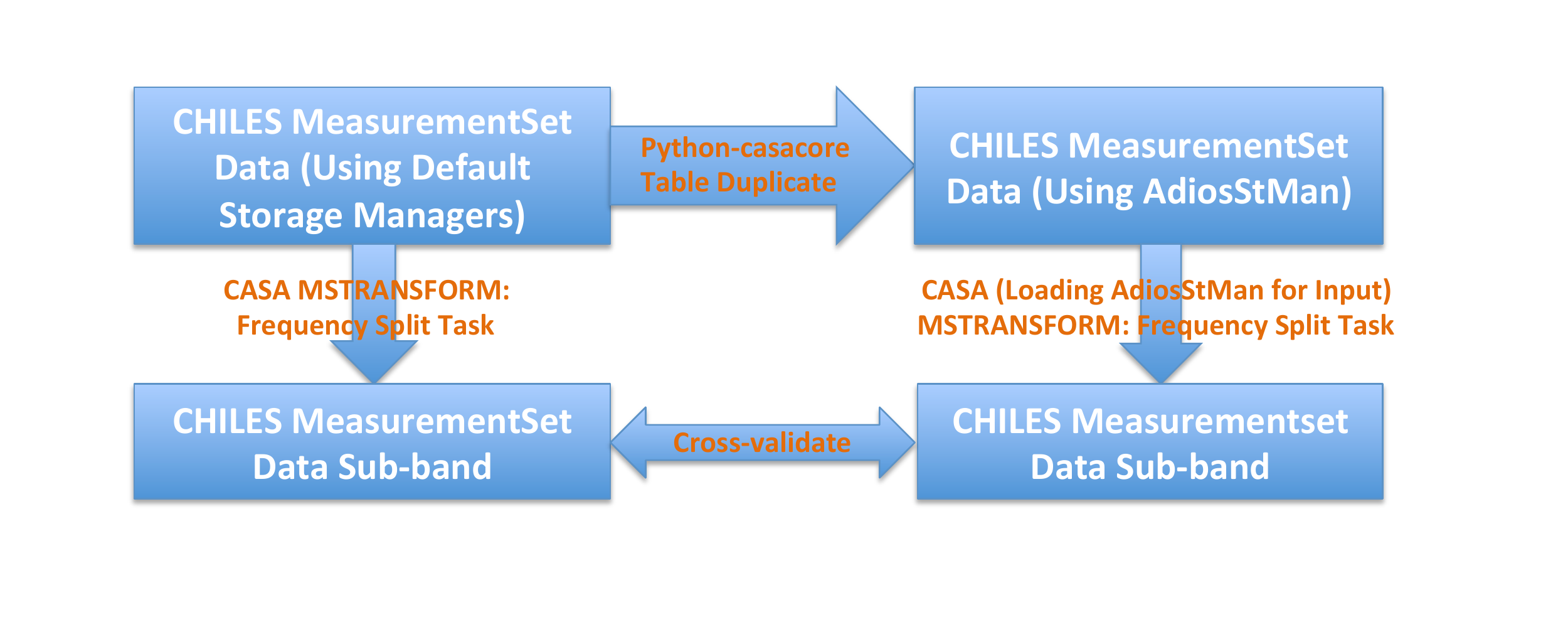}
\caption{
Shown is the work flow of the AdiosStMan validation using the CHILES MeasurementSet data.
}
\label{fig:chiles}
\end{center}
\end{figure}

\begin{figure}
\begin{center}
\includegraphics[width=0.5\textwidth]{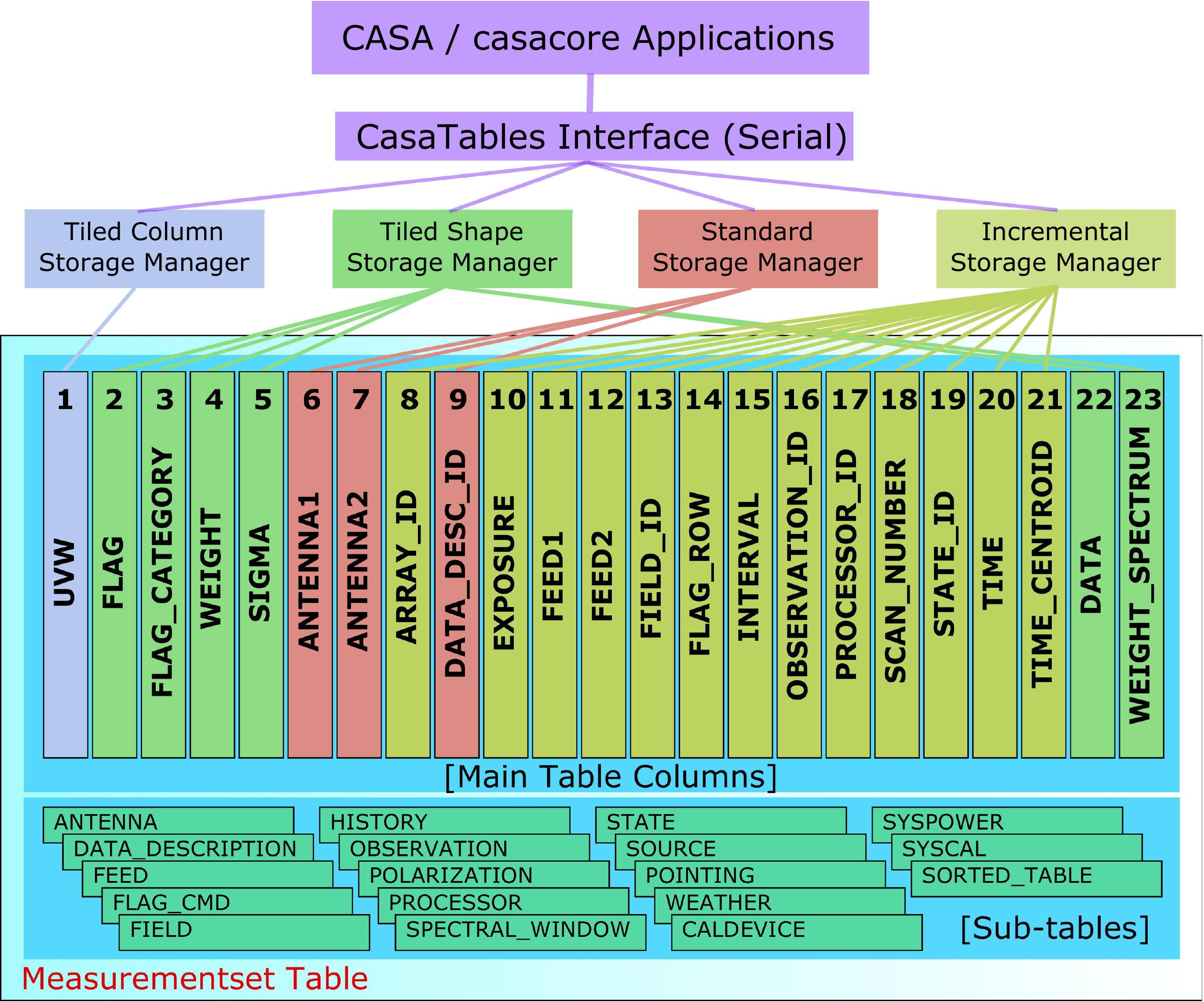}
\caption{
Shown is the structure of a MeasurementSet table from the CHILES project.
}
\label{fig:ms}
\end{center}
\end{figure}

\begin{figure}
\begin{center}
\includegraphics[width=0.75\textwidth]{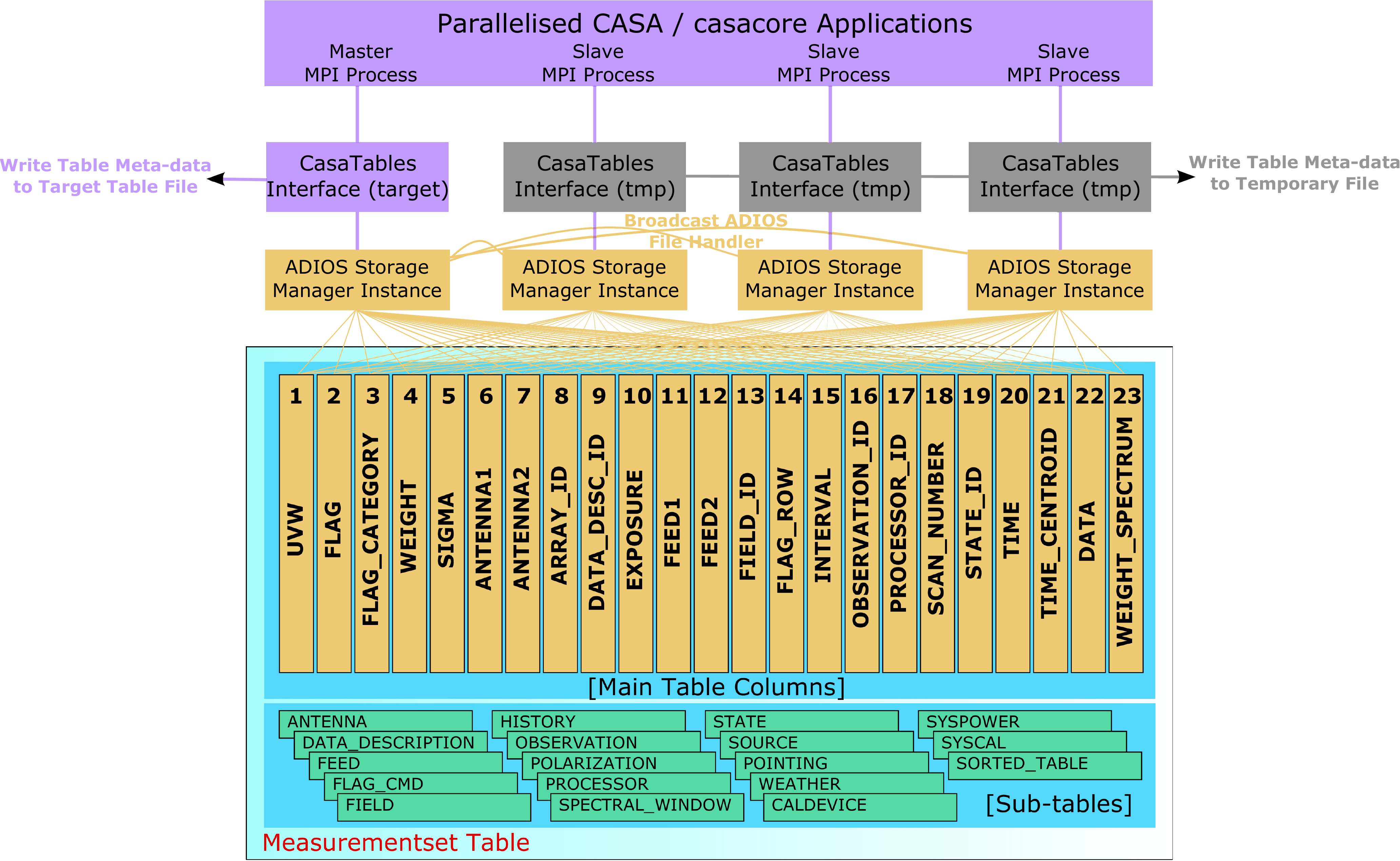}
\caption{
Shown is how the ADIOS Storage Manager accesses the MeasurementSet table given in Figure \ref{fig:ms}.
}
\label{fig:ms_adios}
\end{center}
\end{figure}

A typical use case of the Casacore Table Data System is the MeasurementSet table.
It is generally used to store visibility data or single-dish data.
Readers are directed to \cite{measurementset} for the definition of the MeasurementSet table.
To validate our ADIOS Storage Manager, we took the MeasurementSet tables from the CHILES project \cite{chiles}, converted it using the ADIOS Storage Manager, applied the MSTransform of CASA on both datasets to do the frequency split task, and then verified the results. 
The work flow of this experiment is given in Figure \ref{fig:chiles}.

\subsection{Frequency Split on CHILES MeasurementSet Data}
\label{section:ms}

The CHILES project runs at the Karl G. Jansky Very Large Array (VLA) telescope, with the upgraded receivers and correlator \cite{vla}.
The CHILES MeasurementSet table we used in this work consists of 23 columns, 1,374,749 rows and 18 subtables, as shown in Figure \ref{fig:ms}.
The visibility data is stored in the DATA column, which is essentially a two dimensional complex array with the size 2 * 2048.
This results in a data size of 32768 Byte for each row, and approximately 45 GB for the entire DATA column.
The main table is managed using four casacore storage managers, the standard storage manager, the incremental storage manager, the tiled column storage manager, and the tiled shape storage manager, as each of them is optimized for a certain type of columns with a certain data pattern.
By replacing these storage managers with the ADIOS Storage Manager, each column of the table can be written from multiple processes.
The main table was then converted in a pattern as shown in Figure \ref{fig:ms_adios}, in which all columns are stored using the ADIOS Storage Manager.

The conversion was done in two different ways.
In one approach, we wrote a piece of MPI code in C++ calling the casacore library to duplicate a casacore table.
This code can be executed in parallel under the MPI environment.
In this case, each MPI process writes a subset of the table rows, across all columns.
However, this code was specifically written for the CHILES MeasurementSet tables.
As there are some optional columns defined in the MeasurementSet table specification, MeasurementSet tables from other projects may have a different number of columns with different data pattern.
Our conversion code may not be able to handle these tables correctly.

In the other approach, we wrote a piece Python code calling the python-casacore wrapper \cite{python-casacore} to do the conversion.
As Python is a dynamically typed language, the python-casacore wrapper provides an automatic table duplicate routine, which can handle arbitrary table structures at runtime.
It also reduced the conversion code from thousands of lines to tens of lines, compared with the first approach.
The disadvantage of this approach is that it can only work in serial, since python-casacore does not provide multi-processing interfaces.
The python-casacore wrapper is built upon the official casacore version on Github \cite{casacore}, and therefore, it can be applied directly using a compatible casacore installation with the ADIOS Storage Manager compiled and included.

The second step of this experiment was to apply the MSTransform routine of the CASA package to both the original and the ADIOS MeasurementSet tables to do the frequency split.
The frequency split task essentially reads in a subset of the frequency channels from the input MeasurementSet table, and writes them out to a new MeasurementSet table.
After applying the frequency split task, the final step was to verify the two sub-band MeasurementSet tables.
This was done using a C++ program, which reads in two MeasurementSet tables row by row and compares the data for all columns.

\subsection{Outcomes and Improvements to AdiosStMan}

Until this MeasurementSet experiment, the ADIOS Storage Manager had only been compiled and used with the official casacore, while this MeasurementSet experiment needed a CASA specific version of the ADIOS Storage Manager.
Before casacore Version 2.0.2, the official casacore library was not compatible with CASA's built-in casacore.
Starting from 2.0.2, it has been claimed that casacore can be used to compile CASA.
However, we were not successful in doing so, using the latest CASA 4.5 and casacore 2.0.3 source code.
This means that we were not able to use the AdiosStMan library files that we compiled for the official casacore directly with CASA.
Instead, we needed to re-compile the ADIOS Storage Manager based on CASA's built-in casacore, which differs from the official casacore in some of the interfaces.
After some effort toward this end, the ADIOS Storage Manager has become compatible with CASA for the MSTransform task since Version 0.6.

Due to the nature of ADIOS, an ADIOS array variable can only be defined with a fixed dimension and shape.
This fundamentally limits the ADIOS Storage Manager in a way that it is not feasible to handle non-fixed-shape array columns of the casacore table data system.
However, in reality, most MeasurementSet table generators are implemented to always instantiate array column objects using non-fixed shapes, regardless of the fact that the vast majority of the array columns actually always store fixed-shape arrays.
This unnecessarily makes the ADIOS Storage Manager ineffective in most MeasurementSet table use cases.
To resolve this problem, starting from Version 0.7, we hard-coded the ADIOS Storage Manager to treat any array columns as fixed-shape columns.
This feature can be switched on by defining the FORCE\_DIRECT\_ARRAY macro at compile time.
Once it is enabled, the ADIOS Storage Manager ignores all predefined array shape information.
Instead, it only determines the column shape when the first write operation is conducted, and uses it as the fixed shape for the entire column.

The CASA MSTransform code accesses MeasurementSet tables in a very inefficient way.
It reads a single row at a time, regardless of the fact that most of the use cases actually always process a group of contiguous rows.
This becomes critical while processing a MeasurementSet table containing millions of rows, not only for the reason that accessing files on disks so frequently causes huge overheads, but also because calling the casacore table interface itself on such a basis could be expensive as well.
The built-in storage managers in casacore did an excellent job to optimize the read performance for this type of inefficient access pattern.
The ADIOS Storage Manager had been performing orders of magnitude slower in such use cases, until in Version 0.7, we introduced a cache mechanism to prefetch data for contiguous rows.
Using the latest version of AdiosStMan, we saw similar read performance to casacore built-in storage managers.

The outcome of this experiment indicates that our ADIOS Storage Manager works properly in a complex use case scenario involving casacore, CASA, python-casacore and MeasurementSet tables.
In terms of the performance, it takes approximately 60 seconds to finish the frequency split task on the original MeasurementSet table, and approximately 73 seconds on the AdiosStMan MeasurementSet table.
For both of the two jobs, the input MeasurementSet is approximately 45GB, and the output approximately 381MB.
These numbers were obtained by running the single-threaded CASA MSTransform task on a single node of the Fornax supercomputer of the Pawsey Supercomputing Centre.
The hardware specification of the Fornax supercomputer is given in Section \ref{section:testbed}.
The frequency split task is a combination of read, write, and some lightweight calculations.
Therefore, it is not recommended to use these numbers for deriving any I/O throughput figures.
We will present more adequate performance testing results in upcoming sections.

\section{Parallel I/O Performance Testing}

In this section, we will present some performance testing results for both parallel writing and parallel reading and relevant analyses.
The testbed we used for the work we will describe in this section is the Magnus supercomputer at the Pawsey Supercomputing Centre.
The hardware specification of Magnus is given in~\ref{section:testbed}.
However, by the time we ran these tests, Magnus had undergone a major upgrade.
The number of nodes increased to 1536, each node hosting two 12-core Intel Xeon E5-2690V3 CPUs.
The Lustre storage backend is still as described in~\ref{section:testbed}.

\subsection{Parallel Array Write Test}
\label{section:parallel_write}

The parallel array writing test focuses on the AdiosStMan performance as it is the only CTDS storage manager supporting parallel writing at the physical table level that we are aware of.
Testing was configured to write a casacore table from between 4 and 80 nodes, each node launching one MPI process.
The table consists of an array column, each column cell being a 36 MB floating-point array.
The number of rows of the table varies from 200 to 1000, which results in the table size varying from 7.2 GB to 36 GB\@.
In this test, AdiosStMan was configured to use the MPI\_AGGREGATE (previously called MPI\_AMR) transport method~\cite{adios_lessons} of ADIOS, which provides a two-layer buffering mechanism to optimize operations for small size data.
The Lustre stripe size used in this test is 32.
All testing results were averaged from 20 runs.

\begin{figure}
\begin{center}
\includegraphics[width=0.6\textwidth]{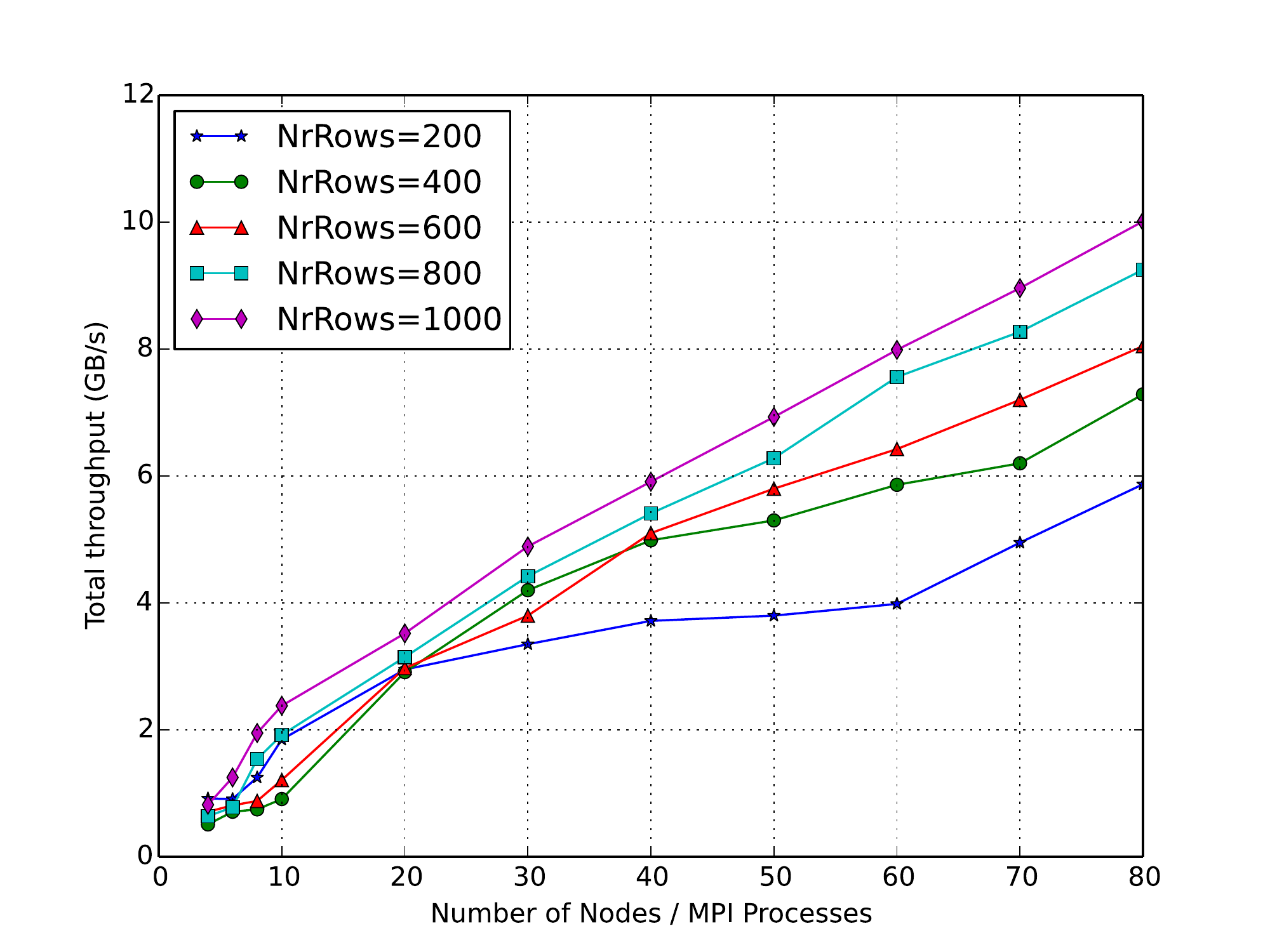}
\caption{Shown are AdiosStMan testing results for parallel array writing.
Testing was configured to write a casacore table from between 4 and 80 nodes on Magnus, each node launching one MPI process.
The table consists of an array column, each column cell being a 36 MB floating-point array.
The number of rows of the table varies from 200 to 1000, which results in the table size varying from 7.2 GB to 36 GB\@.
}
\label{fig:write}
\end{center}
\end{figure}

As shown in Figure~\ref{fig:write}, AdiosStMan basically achieved a linear scale up on up to 80 compute nodes of Magnus, except for cases where the table does not contain enough rows to boost the performance.
The highest throughput we saw is approximately 10 GB/s on 80 nodes, for a 36 GB table consisting of 1000 rows.
This number is close to what is in the ADIOS benchmark results shown in Figure~\ref{fig:bench}, where we saw a 12.5 GB/s throughput on 80 nodes for a 16 GB table, and 15 GB/s for a 9 GB table.

The decrease in performance can be accounted to several reasons.
Firstly, since the ADIOS benchmark was conducted, the compute component of Magnus has been upgraded from 208 nodes to 1536 nodes, while the storage component was untouched.
This potentially affects the per-node bandwidth between compute nodes and storage nodes.
Secondly, the data pattern of the ADIOS benchmark is not identical to that of the AdiosStMan parallel array write test, although similar.
Therefore, the comparison may not be exactly fair.
Thirdly, in the ADIOS benchmark, the ADIOS transport method we used was POSIX\@.
However, while implementing AdiosStMan, we realized that casacore tables written using the MPI\_AGGREGATE method would result in much better read performance.
Therefore, the default transport method in AdiosStMan was set to MPI\_AGGREGATE, and thus it was adopted in this test as well.
We have not systematically quantified the difference between these two transport methods in various writing scenarios.
This, however, could be another factor of the difference between the ADIOS benchmark results and the AdiosStMan testing results.
Finally, for AdiosStMan, calling the CTDS interface may cause extra overhead, which potentially affects the throughput figures.

Similar to the ADIOS benchmark results, the throughput scaling is basically linear but not optimal, which means doubling the compute nodes do not produce twice as much the throughput.
This could be limited by the network topology used to connect the compute nodes.
For example, there might be a physical limitation on the overall bandwidth between the compute nodes allocated and the storage nodes.
Another possible reason is that the ADIOS transport methods used in these tests require MPI communications between compute nodes while writing data.
These MPI communications include collective calls, broadcasts and peer-to-peer messages, which could lead to increasing overheads as the number of compute nodes grows.

\subsection{Array Read Test}
\label{section:parallel_read}
\begin{figure}
\begin{center}
\includegraphics[width=0.6\textwidth]{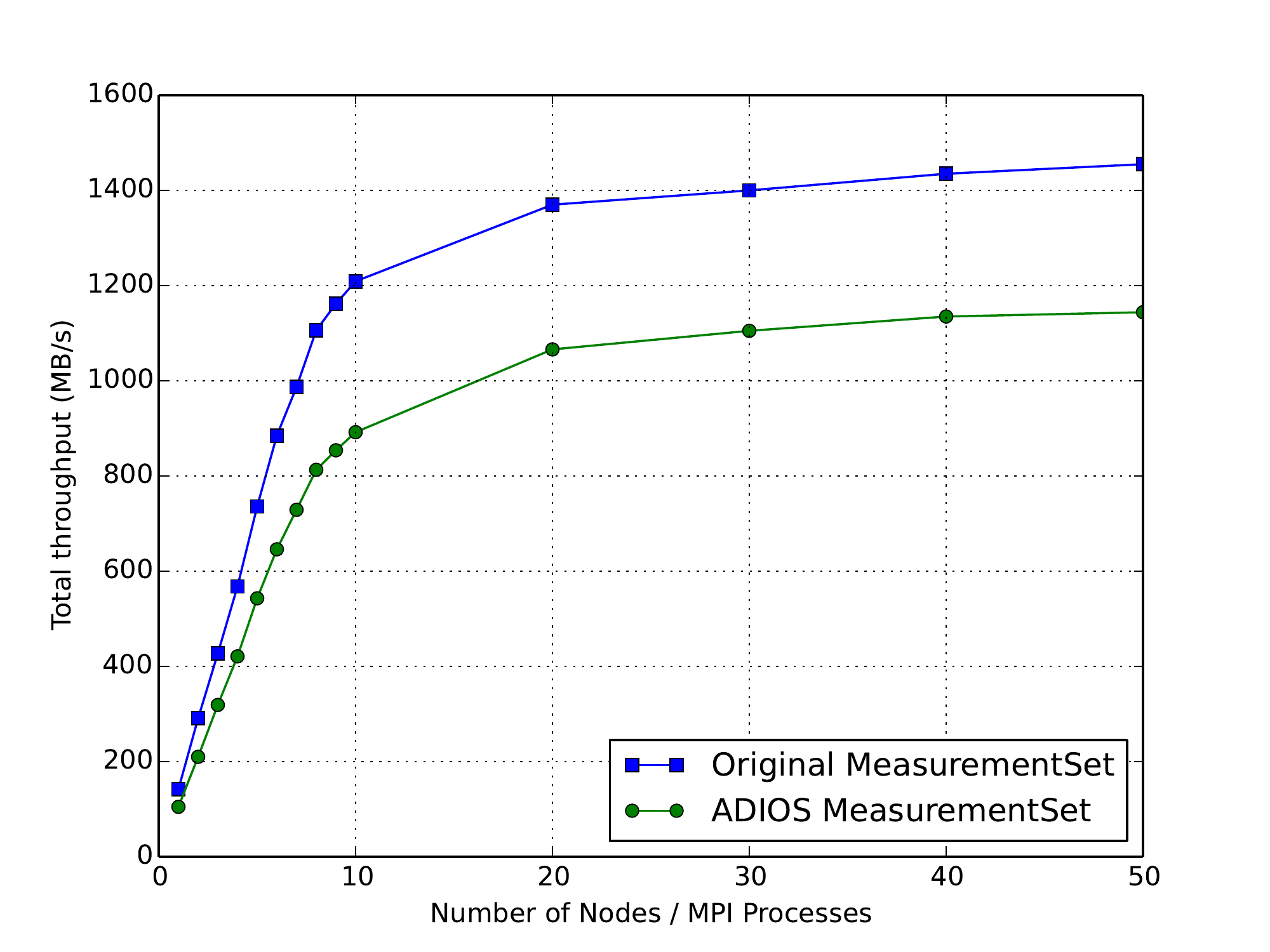}
\caption{Shown are AdiosStMan testing results for parallel array reading. The total throughput is measured in MB/s.
Testing was schemed to read a MeasurementSet table from between 1 and 50 nodes, each node launching one MPI process.
Each process reads the entire DATA column, which is approximately 45 GB, into memory on a node independent of other processes.
}
\label{fig:read_mbps}
\end{center}
\end{figure}

\begin{figure}
\begin{center}
\includegraphics[width=0.6\textwidth]{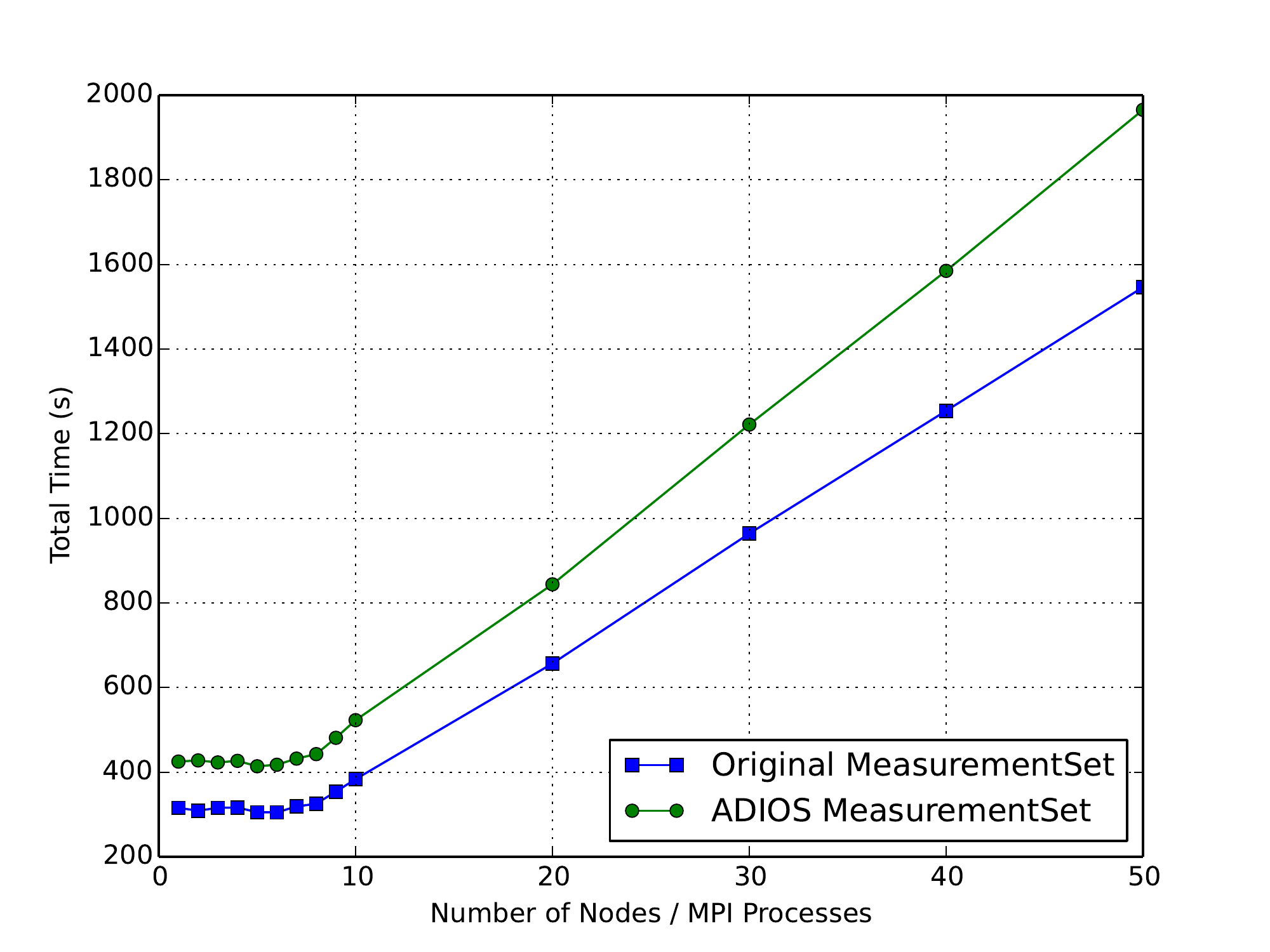}
\caption{Shown are the same AdiosStMan testing results for parallel array reading as Figure~\ref{fig:read_mbps}, but presented are execution time in seconds.
}
\label{fig:read_sec}
\end{center}
\end{figure}

The array reading test compares AdiosStMan and TiledShapeStMan, since the Casacore Table Data System and its built-in storage managers allow a table to be accessed from multiple processes for reading.
Testing was schemed to read a single MeasurementSet table from between 1 and 50 nodes, each node launching one process.
All processes were launched at the same time using mpirun.
Each process reads the entire DATA column into its local memory, which is approximately 45 GB\@.
The MeasurementSet tables we used in this test were from the MeasurementSet experiment described in Section~\ref{section:ms}.
The original CHILES MeasurementSet table uses the Tiled Shape Storage Manager for the DATA column, while the ADIOS MeasurementSet uses the ADIOS Storage Manager.
Performance was measured for both total throughput in megabytes per second (Figure~\ref{fig:read_mbps}) and execution time in seconds (Figure~\ref{fig:read_sec}).
The Lustre stripe size for the testing data files was 8.

A fundamental difference between parallel reading and parallel writing on the Lustre filesystem is that for writing, data from different writers can usually be distributed as scattered as configurable onto different object storage targets.
This helps the overall throughput more easily scale up linearly with the number of writers, or compute nodes.
However, for parallel reading, the files to read are already in place on some object storage targets.
It is much more likely that all readers try to access the same object storage target at the same time.
Therefore, we see a much worse scalability of parallel reading than parallel writing.
More specifically, it is noticeable in Figure~\ref{fig:read_mbps} that the overall throughput almost flats out after 20 compute nodes, whereas for parallel writing it scales nearly linearly up to 80 nodes.

It also should be mentioned that AdiosStMan performs 20\% to 30\% slower than TiledShapeStMan.
The reason behind may be because ADIOS is mainly optimized for use cases where the number of readers are identical or close to the number of writers.
However, the ADIOS MeasurementSet table we used in this test was written from only one process using the python-casacore conversion code mentioned in Section~\ref{section:ms}, which does not match the number of readers in this test.
It makes a fair comparison though, because the testing data for TiledShapeStMan was also written from a single process.

Considering these factors, this test does not intend to demonstrate parallel reading performance of either storage manager, but rather to validate if AdiosStMan works properly for reading in a real-world use case, as a supplementary to the MeasurementSet experiment.
For a sensible parallel reading test scheme, we found it difficult to make in a convincing and practical way at current stage.
The data cannot be read in a too regular manner, which a testing program usually is.
Because in this case, all processes read the same partition of a file, which causes a serious bottleneck.
Thus, the I/O throughput of the Lustre filesystem and its cache system has overly large impacts on the result, compared to the performance of the storage manager itself.
It is difficult to test using a real-world user application too, since this type of massively parallel data I/O has hardly been used in real-world data processing pipelines.
If there is a candidate use case, it then involves separating out the I/O component to ensure it is the only resource consumer, which could be an enormous amount of work.
Moreover, it still may not be convincing as it is only one of the many access patterns.
A feasible way could be to abstract some most used data reading patterns, develop benchmark applications accordingly, and carry out a more systematic test and analysis, which is left to future work.

\section{Conclusion}

In this paper we have presented a parallel Casacore Table Data System storage manager using the Adaptive IO System as the storage backend.
The decision of using ADIOS was made based on the results of an elaborate benchmark comparing the parallel I/O performance between HDF5 and ADIOS.
The ADIOS Storage Manager proved to work properly with the MeasurementSet, which is one of the typical use cases of CTDS.
In the performance test, AdiosStMan showed an excellent scalability on up to 80 compute nodes for parallel writing, while other storage managers were not taken into the test as they are not able to handle parallel writing.
For parallel reading, limited by the nature of the Lustre filesystem, neither of AdiosStMan and TiledShapeStMan scaled well, and AdiosStMan was 20\% to 30\% behind TiledShapeStMan.

\subsection{Future Work}

As shown and discussed in Section \ref{section:parallel_read}, AdiosStMan fell behind TiledShapeStMan by 20\% to 30\% in parallel reading.
This indicates that there is still considerable space for AdiosStMan to be optimized in this aspect.
One of the recommended ways of doing so is to experiment with different ADIOS transport methods and / or configuration parameters.
Also, a more systematic parallel reading test can help understand the I/O behavior of AdiosStMan, for further optimization.

The latest ADIOS release, Version 1.9.0, provides a new feature that enables the POSIX transport method to open an existing ADIOS file for appending data.
This can effectively solve the problem of the POSIX transport method that it requires the cache size of the entire ADIOS group even though every node only handles a small proportion of the group.
Taking advantage of this feature, a CTDS table will be able to store as many rows as the filesystem allows in a single file, rather than only no larger than the physical memory of a compute node previously.
Furthermore, using POSIX as the main transport method of AdiosStMan may also solve the MPI communication overhead problem mentioned in Section \ref{section:parallel_write}, because it requires far less inter-node communications than other transport methods.

One of the obstacles preventing AdiosStMan from being applied universally is that it requires user applications to change their code, although only a few lines each.
It also means introducing the MPI environment as well, as ADIOS is dependent upon MPI.
By using MPI, it adds collective function calls into user applications.
This will potentially cause problems where there only needs loose synchronizations or even no synchronization at all.
In some extreme cases, this could cause the user application frozen on a collective call.
To perfectly solve this problem, there needs a storage backend which is not dependent on MPI while still achieving similar performance.

From the opposite point of view, since CASA has recently introduced MPI support, it is worth serious investigation to identify a scenario where AdiosStMan can be combined into its concatenate table model.
This is promising in both improving the universality of AdiosStMan, and improving the concatenate table model so that it can achieve parallel writing at the physical table layer.

While AdiosStMan achieves parallel writing by bypassing the CTDS locking mechanism in a non-standard way, CTDS interface is still part of the dataflow.
As a result, some use cases are still limited to serial operations, for instance, when a table is opened for both reading and writing.
To make CTDS a completely parallel system, so far the only solution we could think of is to replace the entire CTDS interface, instead of only investigating the storage manager layer.

\section*{Acknowledgements}
We wish to thank Scott Klasky, Norbert Podhorszki, Jeremy Logan and Qing Liu from the Oak Ridge National Laboratory for suggestions on optimizing this work.

This work was supported by resources provided by the Pawsey Supercomputing Centre with funding from the Australian Government and the Government of Western Australia.

The Karl G. Jansky Very Large Array and the National Radio Astronomy Observatory is a facility of the National Science Foundation operated under cooperative agreement with Associated Universities Inc.. We wish to thank the CHILES team for flagging and calibrating the data used in this work.

\bibliographystyle{elsarticle-num}
\bibliography{mybibtex}

\end{document}